\documentclass[11pt,a4paper]{article}
\usepackage{amssymb} \usepackage{amsmath} \usepackage{graphicx}
\usepackage{epsfig,latexsym}
\begin{document}

\def\bef{\begin{figure}}
\def\eef{\end{figure}}
\newcommand{\ans}{ansatz }
\newcommand{\be}[1]{\begin{equation}\label{#1}}
\newcommand{\beq}{\begin{equation}}
\newcommand{\ee}{\end{equation}}
\newcommand{\beqn}[1]{\begin{eqnarray}\label{#1}}
\newcommand{\eeqn}{\end{eqnarray}}
\newcommand{\bd}{\begin{displaymath}}
\newcommand{\ed}{\end{displaymath}}
\newcommand{\mat}[4]{\left(\begin{array}{cc}{#1}&{#2}\\{#3}&{#4}
\end{array}\right)}
\newcommand{\matr}[9]{\left(\begin{array}{ccc}{#1}&{#2}&{#3}\\
{#4}&{#5}&{#6}\\{#7}&{#8}&{#9}\end{array}\right)}
\newcommand{\matrr}[6]{\left(\begin{array}{cc}{#1}&{#2}\\
{#3}&{#4}\\{#5}&{#6}\end{array}\right)}
\newcommand{\cvb}[3]{#1^{#2}_{#3}}
\def\lsim{\raise0.3ex\hbox{$\;<$\kern-0.75em\raise-1.1ex
e\hbox{$\sim\;$}}}
\def\gsim{\raise0.3ex\hbox{$\;>$\kern-0.75em\raise-1.1ex
\hbox{$\sim\;$}}}
\def\abs#1{\left| #1\right|}
\def\simlt{\mathrel{\lower2.5pt\vbox{\lineskip=0pt\baselineskip=0pt
           \hbox{$<$}\hbox{$\sim$}}}}
\def\simgt{\mathrel{\lower2.5pt\vbox{\lineskip=0pt\baselineskip=0pt
           \hbox{$>$}\hbox{$\sim$}}}}
\def\unity{{\hbox{1\kern-.8mm l}}}
\newcommand{\eps}{\varepsilon}
\def\ep{\epsilon}
\def\ga{\gamma}
\def\Ga{\Gamma}
\def\om{\omega}
\def\omp{{\omega^\prime}}
\def\Om{\Omega}
\def\la{\lambda}
\def\La{\Lambda}
\def\al{\alpha}
\newcommand{\ov}{\overline}
\renewcommand{\to}{\rightarrow}
\renewcommand{\vec}[1]{\mathbf{#1}}
\newcommand{\vect}[1]{\mbox{\boldmath$#1$}}
\def\tm{{\widetilde{m}}}
\def\mcirc{{\stackrel{o}{m}}}
\newcommand{\Dm}{\Delta m}
\newcommand{\dm}{\varepsilon}
\newcommand{\tanb}{\tan\beta}
\newcommand{\nbar}{\tilde{n}}
\newcommand\PM[1]{\begin{pmatrix}#1\end{pmatrix}}
\newcommand{\up}{\uparrow}
\newcommand{\down}{\downarrow}
\def\omE{\omega_{\rm Ter}}
%

\newcommand{\Dsusy}{{susy \hspace{-9.4pt} \slash}\;}
\newcommand{\DCP}{{CP \hspace{-7.4pt} \slash}\;}
\newcommand{\mc}{\mathcal}
\newcommand{\gr}{\mathbf}
\renewcommand{\to}{\rightarrow}
\newcommand{\gtc}{\mathfrak}
\newcommand{\wh}{\widehat}
\newcommand{\br}{\langle}
\newcommand{\kt}{\rangle}


\def\lsim{\mathrel{\mathop  {\hbox{\lower0.5ex\hbox{$\sim$}
\kern-0.8em\lower-0.7ex\hbox{$<$}}}}}
\def\gsim{\mathrel{\mathop  {\hbox{\lower0.5ex\hbox{$\sim$}
\kern-0.8em\lower-0.7ex\hbox{$>$}}}}}

\def\nn{\\  \nonumber}
\def\de{\partial}
\def\brf{{\mathbf f}}
\def\bbf{\bar{\bf f}}
\def\bF{{\bf F}}
\def\bbF{\bar{\bf F}}
\def\bA{{\mathbf A}}
\def\bB{{\mathbf B}}
\def\bG{{\mathbf G}}
\def\bI{{\mathbf I}}
\def\bM{{\mathbf M}}
\def\bY{{\mathbf Y}}
\def\bX{{\mathbf X}}
\def\bS{{\mathbf S}}
\def\bb{{\mathbf b}}
\def\bh{{\mathbf h}}
\def\bg{{\mathbf g}}
\def\bla{{\mathbf \la}}
\def\bmu{\mathbf m }
\def\by{{\mathbf y}}
\def\bmu{\mbox{\boldmath $\mu$} }
\def\bsig{\mbox{\boldmath $\sigma$} }
\def\bunity{{\mathbf 1}}
\def\cA{{\cal A}}
\def\cB{{\cal B}}
\def\cC{{\cal C}}
\def\cD{{\cal D}}
\def\cF{{\cal F}}
\def\cG{{\cal G}}
\def\cH{{\cal H}}
\def\cI{{\cal I}}
\def\cL{{\cal L}}
\def\cN{{\cal N}}
\def\cM{{\cal M}}
\def\cO{{\cal O}}
\def\cR{{\cal R}}
\def\cS{{\cal S}}
\def\cT{{\cal T}}
\def\eV{{\rm eV}}
%


\begin{flushright}
ROM2F/2015/03\\
\end{flushright}
\vspace{0.5cm}

\large
 \begin{center}
 {\Large \bf Neutron Majorana mass from Exotic Instantons \\
 in a Pati-Salam model}
 \end{center}

 \vspace{0.1cm}

 \vspace{0.1cm}
 \begin{center}
{\large Andrea Addazi}\footnote{E-mail: \,  andrea.addazi@infn.lngs.it} \\
{\it \it Dipartimento di Fisica,
 Universit\`a di L'Aquila, 67010 Coppito, AQ \\
LNGS, Laboratori Nazionali del Gran Sasso, 67010 Assergi AQ, Italy}
\end{center}

 \begin{center}
{\large Massimo Bianchi}\footnote{E-mail: \, massimo.bianchi@roma2.infn.it}
\\
{\it Dipartimento di Fisica, Universit\`a di Roma ÒTor VergataÓ, \\
I.N.F.N. Sezione di Roma ÒTor VergataÓ, \\
Via della Ricerca Scientifica, 1 00133 Roma, ITALY}
\end{center}

\vspace{1cm}
\begin{abstract}
\large

We show how exotic stringy instantons can generate 
an effective interaction between color diquark sextets 
 in a Pati-Salam model, 
inducing a Majorana mass term for the neutron.
In particular, we discuss a simple quiver theory
for a Pati-Salam like model, 
as an example in which the calculations
of exotic instantons' effects are simple
and controllable. 
We discuss some different possibilities in order to 
generate $n-\bar{n}$ oscillation testable in the next generation of experiments, 
Majorana mass matrices for neutrini and a Post-Sphaleron Baryogenesis scenario. 
Connections with Dark Matter issues and the Higgs mass Hierarchy problem are discussed, 
in view of implications for 
 LHC and rare processes physics.
The model may be viewed as a completion of Left-Right symmetry, 
alternative to a GUT-inspired scenario.
Combined measures in Neutron-Antineutron physics, FCNC, LHC, Dark Matter 
could rule out the proposed model or uncover aspects of physics at the Planck scale!

\end{abstract}

\baselineskip = 20pt

\section{Introduction}
 
How can Matter be generated in our Universe?
And how are neutrino masses generated? 
Has the neutron a Majorana mass?

In principle, these three questions could appear unrelated.
However, in Left-Right symmetric models with $SU(2)_{L}\times SU(2)_{R}\times SU_{c}(3) \times U(1)_{B-L}$ gauge group, one can find intriguing and elegant connections between 
these three issues. A Left-Right model is naturally embedded in a
Pati-Salam (P-S) model with $G_{224}=SU(2)_{L}\times SU(2)_{R}\times SU(4)_{c}$ gauge group \cite{PS}, 
that in turn can be embedded in an $SO(10)$ GUT.

As originally suggested in \cite{MM}, new Higgses $\Delta_{R}$ in the $({\bf 1,3,10})$ (and $\Delta_{L}$ in the $({\bf 3,1,10^*})$) of $G_{224}$ can be introduced in PS models
in order to spontaneously break Left-Right symmetry, through 
$\langle \Delta_{R}\rangle=v_{R}\neq 0$ and $\langle\Delta_{L}\rangle=0$. 
This mechanism also produces Majorana masses for Right-Handed neutrinos,  that can trigger a seesaw mechanism as suggested in \cite{MohaSenj}, 
spontaneously breaking $U(1)_{B-L}$ at the same time. 
The new Higgs $\Delta_{R}(1,3,10) \equiv \Delta^{c}(1,3,10^*)$ decomposes with respect to $SU(2)_{L}\times SU(2)_{R}\times SU(3)_{C}\times U(1)_{B-L}$ as
\be{DeltaR}
\Delta_{R}(1,3,10)=\left\{(1,3,1)_{-2}+(1,3,3)_{-2/3},+(1,3,6)_{2/3}\right\}_{R}
\ee
with $\Delta^{c}_{l^{c}l^{c}}(1,3,1)_{-2}$ 
generating Right-Handed neutrini masses via
$\langle\Delta^{c}_{\nu^{c}\nu^{c}}\rangle\nu^{c}\nu^{c}$.  

In GUT $SO(10)$, the $({\bf 1,3,10})$ of $G_{224}$ and its conjugate are contained in the ${\bf 126}$ representation\footnote{The complete decomposition reads ${\bf 126}\rightarrow 
({\bf 1,3,10}) + ({\bf 3,1,10^*}) + ({\bf 2,2,15}) + ({\bf 1,1,6})$.}. 
But $\Delta^{c}(1,3,10)$ also contains {\it color sextet diquark fields} $\Delta^{c}_{q^{c}q^{c}}(1,3,6)_{2/3}$,
leading to possible new effects. 
In particular, these sextets can induce Baryon number violating
processes beyond the Standard Model (BSM). 
Color sextets can also play an important role in some post-sphaleron baryogenesis mechanism
\cite{BM1,BM2,BM3}. In susy extensions, a quartic superpotential term 
$$
{\mathcal W}_4 = \Delta^{c} \Delta^{c} \Delta^{c} \Delta^{c}/\mathcal{M}_{0}
$$ 
can appear that, among other terms, produces a term coupling three color sextets $\Delta^{c}_{q^{c}q^{c}}$ and one color singlet  
$\Delta^{c}_{\nu^{c}\nu^{c}}$, 
as $\Delta^{c}_{u^{c}u^{c}}\Delta^{c}_{d^{c}d^{c}}\Delta^{c}_{d^{c}d^{c}}\Delta_{\nu^{c}\nu^{c}}$.
When the color singlet $\Delta^{c}_{\nu^{c}\nu^{c}}$ 
takes an expectation value, $U(1)_{B-L}$ is spontaneously broken
and Right-handed neutrini get a mass \cite{MohaSenj}. 
Moreover a Majorana mass for the neutron is generated through the processes shown in Fig.~1-(a)-(b).
This can be directly tested in Neutron-Antineutron transition experiments, as firstly proposed in \cite{MM}. 
As shown in \cite{BFM}, constraints from post-sphaleron baryogenesis, 
and neutrino oscillations imply a precise 
prediction about neutron-antineutron transitions: 
an oscillation time $\tau_{n-\bar{n}} \approx 10^{10} s$ accessible to the 
next generation of experiments!

In principle, color sextet scalars could be as light as $1\, \rm TeV$ and they could be 
directly searched at the LHC, as proposed in \cite{LHC}: dijet data put constraints 
on the couplings between colored scalars and quarks. In \cite{LHCexp}, bounds 
are shown in comparison with LHC data. 
On the other hand, FCNC processes could impose stronger constraints 
on the sextets with respect to LHC direct searches (see \cite{FCNC} for comparison with experimental limits). 
For example, the $\Delta^{c}_{dd}$ field couples to two down-type quarks $dd, ss, bb$: it mediates $B^{0}_{d,s}-\bar{B}^{0}_{s,s}$, $K^{0}-\bar{K}^{0}$ oscillations and 
$B$ mesons decays. On the other hand $\Delta^{c}_{uu}$ mediates $D^{0}-\bar{D}^{0}$ oscillations and $D$-decays like $D\rightarrow K\pi, \pi\pi$. 
These analyses show that for coupling strengths of order $10^{-2}$, 
the mass of the color sextets has to exceed the TeV-scale.

\begin{figure}[t]
\centerline{ \includegraphics [height=8cm,width=1.2 \columnwidth]{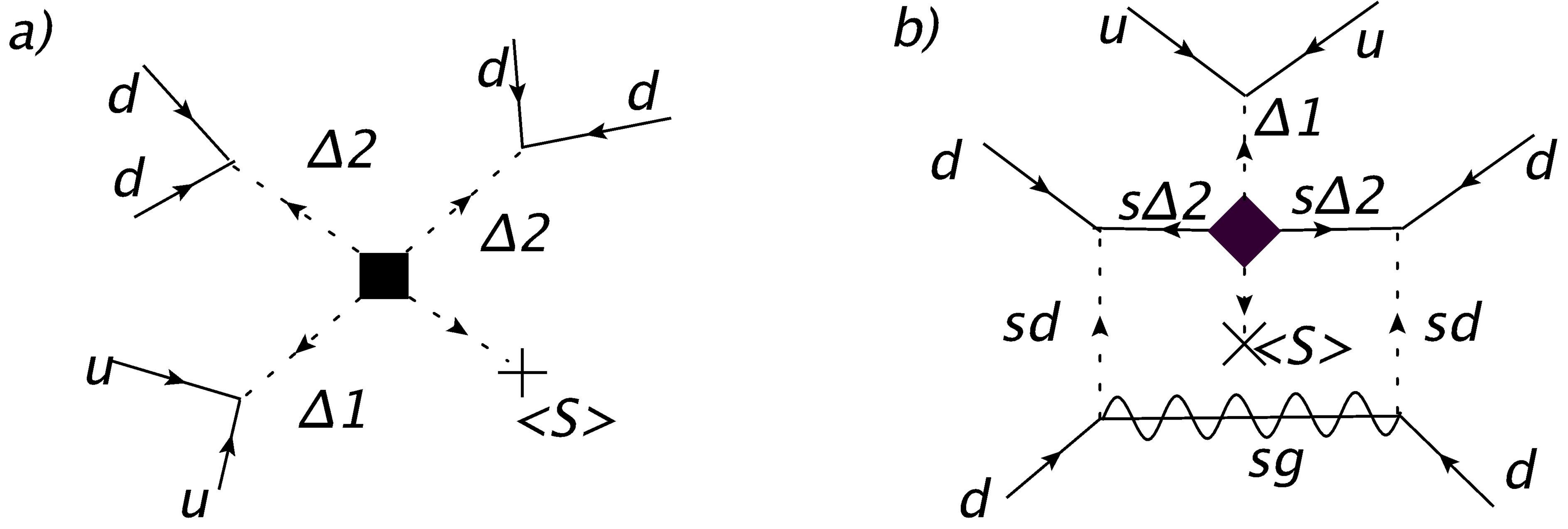}}
\vspace*{-1ex}
\caption{ a) We show the main diagram for neutron-antineutron oscillation in a non-susy $SU(4)_{c}\times SU(2)_{L}\times SU(2)_{R}$ model \cite{Md}. The transition is induced by color sextets $\Delta^{c}_{u^{c}u^{c}}$ and $\Delta^{c}_{d^{c}d^{c}}$. b) We show the main diagram for neutron-antineutron oscillation in a supersymmetric $SU(4)_{c}\times SU(2)_{L}\times SU(2)_{R}$ model \cite{Md}. The transition is induced by color sextets $\Delta^{c}_{u^{c}u^{c}}$ and $\Delta^{c}_{d^{c}d^{c}}$ arising from the decomposition of $\Delta^{c}(1,3,10)$. The latter participate in a non-perturbative quartic superpotential term. 
The diagram involves also gaugini $\tilde{g}$ (gluini, zino or bino), squarks $\tilde{d}^{c}$, and susy partners of the color sextets $\tilde{\Delta}^{c}_{d^{c}d^{c}}$.  }
\label{plot}   
\end{figure}

In this paper, we propose a (susy-)PS model that is alternative to the $SO(10)$ GUT inspired
model mentioned above. We consider an (un)oriented open string model with intersecting 
D-brane stacks, producing a susy PS like model. Models of this kind have been previously considered {\it e.g.} in \cite{Anastasopoulos:2010ca}, where an analysis of the mass spectrum and low-energy phenomenology has been carried out. 
In  oriented string theory, a simple way to generate a $U(N)$ gauge theory 
 is to consider a stack of $N$ D-branes, parallel to each other. 
 In this way the excitations of
the open strings stretching between the $N$ D-branes
reproduce at low energy the fields in the adjoint of the $U(N)$ gauge symmetry. 
In type IIA, compactified on a six-dimensional (CY) manifold, 
one can consider stacks of intersecting D6-branes, filling the 4D ordinary Minkowski spacetime, and wrapping internal 3-cycles. 
From strings connecting different stacks of branes, we can construct chiral fermions,
localised at the four-dimensional intersections of two stacks of D6-branes $a$ and $b$, in the 
bi-fundamental representation of $U(N_{a} )\times U(N_{b})$ \cite{sessantadue}.
The net (positive - negative) number of intersections  of two branes $a$ and $b$ is a topological invariant, 
representing the number of massless fermions. In the case in which D-branes are space-time filling, $\Omega$-planes have to be introduced in order to cancel tadpoles and irreducible anomalies \cite{Bianchi:1990yu, Bianchi:1990tb, Bianchi:1991eu, sessantatre, sessantaquattro, MBJFM}. An $\Omega$-plane implements a combination of world-sheet parity and a (non) geometric mirror-like involution in the target space. As a consequence Left- and Right-moving modes of the closed strings are identified; closed and open strings become un-oriented.
More choices for the gauge groups and their representations are allowed \cite{Bianchi:1990yu, Bianchi:1990tb, Bianchi:1991eu}. 
In this way, one can produce 
stacks supporting $U(N)$, $SO(N)$
or $Sp(2N)$ gauge groups.
This is interesting in order to construct realistic gauge groups, with chiral matter in a globally consistent model \cite{Angelantonj:1996uy, Angelantonj:1996mw}.
The closed strings propagate in the entire ten dimensional space-time: some mediate gravitational interactions, some behave as axions or scalar moduli fields. 

In principle, one can construct a PS like 
gauge group $U(4)\times Sp_{R}(2)\times Sp_{L}(2)$ 
or $U(4)\times U_{R}(2)\times U_{L}(2)$
in terms of intersecting D-brane stacks and $\Omega$-planes.
In \cite{Anastasopoulos:2010ca} the case $U(4)\times U_{R}(2)\times U_{L}(2)$ was analysed in some detail.
In the present paper, we focus on the  $U(4)\times Sp_{R}(2)\times Sp_{L}(2)$ case
with an $\Omega^{+}$-plane that requires a stack of  four D-branes and its mirror image under $\Omega$, producing $U(4)$, and two stacks of two D-branes each, identified with their own images under $\Omega$, producing $Sp_{L}(2)$ and $Sp_{R}(2)$.

This model has extra anomalous $U(1)$'s that could seem dangerous from a gauge theory point of view. On the other hand, in string theory, Generalized Chern-Simon (GCS) terms appear that cancel anomalies \cite{Anastasopoulos:2006cz, 216}, in combination with
a generalised Green-Schwarz mechanism \cite{217,218}.
The extra $Z'$ gauge bosons can get a mass through a St\"uckelberg mechanism 
\cite{212,213,214,215}. We will return onto phenomenological implications of this
in the next section.
There is however a real problem in this scenario. It is not possible to represent $(1,3,10)$ in terms of open strings. Perturbative open strings have two ends and can at most carry fundamental charges with respect to two classical gauge groups. On the other hand, the triplet is the adjoint of $Sp(2)$, {\it i.e.} the symmetric product of two doublets, and the decaplet of $SU(4)$ is the symmetric product of two tetraplets.
States in the $(1,3,10)$ (or its conjugate) would correspond to multi-pronged strings with two ends on the $U(4)$ stack and two ends on the $Sp(2)$ stack\footnote{{\bf 3} is also the vector of $SO(3)$ but this would prevent the existence of doublets, which are spinors.} that do not admit a perturbative description. 

On the other hand, we will show that a spontaneously breaking pattern to the SM, giving masses to the neutrini,  can be recovered in this model. In fact, we will see that $\phi_{RR}(1,3,1)$ and $\phi_{LL}(1,1,3)$ appear 
as excitations of open strings with both ends attached to
$Sp(2)_{R}$ or $Sp(2)_{L}$, while $\Delta(1,1,10)$ and its conjugate $\Delta^c(1,1,10^*)$ appear from open strings joining $U(4)$ and $U'(4)$
identified which one other under $\Omega$. 
As a consequence the breaking $U(4)\times Sp_{R}(2) \times Sp_{L}(2) \rightarrow 
SU(3)\times SU_{L}(2)\times U(1)$ is not realized through 
$(1,3,10)$, but through $\phi_{RR}(1,3,1)$ and $\Delta(1,1,10)$\footnote{In \cite{Anastasopoulos:2010ca} breaking triggered by Higgses in the $(1,2,4^*)$ was considered together with mass terms generated by exotic instantons.}:

i) Left-Right symmetry breaking through the expectation values $\langle\phi_{RR}\rangle=v_{R}$
and $\langle\phi_{LL}\rangle=0$;

ii) $U(1)_{B-L}$ symmetry breaking though the expectation of value $\langle S\rangle=v_{B-L}$, 
with $S$ the color singlet contained in $(1,1,10)$. Alternatively, $U(1)_{B-L}$ can be 
broken dynamically by exotic instantons or spontaneously by the compactification.

Similarly to the case of $\Delta(1,3,10)$, color sextets are contained in $\Delta(1,1,10)$. 

Our main suggestion is that the super-potential
\be{ope1}
\mathcal{W}_{eff}=S\Delta_{6} \Delta_{6} \Delta_{6}/\mathcal{M}_{0}
\ee
can be generated by non-perturbative quantum gravity effects peculiar to string theory, called
``exotic instantons''. These are associated to Euclidean branes ($E2$-branes in our case), wrapping internal 3-cycles, that could directly produce such interactions, 
in a calculable and controllable way in models like type IIA (un)oriented strings. 
We would like to stress that this class of instantons exists in string theory only, not in gauge theories. The resulting superpotential term is suppressed by the scale 
$\mathcal{M}_{0}=M_{S}e^{+S_{E2}}$, where $M_{S}$ is the string scale
and $e^{+S_{E2}}$ depends on the `size' of the 3-cycles wrapped by 
the relevant $E2$-brane. 
We would like to remark that 
the suppression scale is higher (in principle also much higher) than the string scale.
This is a peculiarity of the non-renormalizable nature of such a non-perturbative term in the string effective action.
As a consequence, the hierarchy depends on the particular model:
$e^{+S_{E2}}$ can be approximately $1$ for a `small' 3-cycles or $e^{+S_{E2}}>>1$ for a `large' 3-cycles.  
So, depending on the String scale, assumed to be larger than some TeV's at least, and the size of the 3-cycle, it is possible to generate such an operator near the LHC scale or at a much higher scale.
This leads to two very different branches for phenomenology. 
In particular, for $\mathcal{M}_{0}\simeq 10^{13}\, \rm GeV$,
color sextets appear near the TeV scale, 
with potential implications in meson physics and at LHC,
as mentioned above. 
On the other hand, for $\mathcal{M}_{0}\simeq M_{S}\simeq  10\, \rm TeV$,
a post-sphaleron scenario is possible and
testable at the next generation of experiments on neutron-antineutron 
oscillations, with heavy color sextets, at a scale
$m_6 >>TeV$ that can be generated by closed-string fluxes, as shown in \cite{Bianchi:2014qma} for quiver theories and reviewed below. In this case, there is no possibility to
produce the sextets at LHC, and 
FCNC's in the meson sector are strongly suppressed. 
On the other hand, extra $Z'$ at the TeV scale naturally appears in this scenario.
Another relevant and peculiar possibility is a Left-Right breaking scale at TeV, 
compatible with neutron-antineutron physics and Post-Sphaleron scenario.
This is not possible in a $SO(10)$ scenario, as remarked in \cite{BFM}. 
Our string-inspired scenario also naturally provides several candidates of WIMP dark matter as we will see. 

We would like to mention that such an operator as (\ref{ope1})
can emerge from stringy dynamics also 
in other kinds of models like F-theory \cite{41,42,43,44,45,46,47}, $E_{8}\times E_{8}$ and $SO(32)$  heterotic strings \cite{32,33,34,35,36,37,38,39,40}, generating an $SO(10)$ GUT. 
For example in heterotic string theories {\it world-sheet instantons}, suppressed by $e^{-R^2/\alpha^\prime}$ and thus perturbative in the string coupling $g_s$,
can induce non-vanishing couplings of the desired kind from such amplitudes 
as $\langle V_{\bf 126} V_{\bf 126}V_{\bf 126}V_{\bf 126}\rangle$, 
for vertex operators $V_{\bf 126}$ that can appear in twisted sectors. 
Unfortunately, in the F-theory case the calculations are more involved \cite{Bianchi:2011qh, BianchiF}.

The paper is organized as follows: 
in Section 2, we briefly review what are stringy instantons and quivers.
In Section 3 we propose a simple and consistent quiver for a Pati-Salam model
generating a Majorana mass for the neutron through exotic instantons. 
In Section 4 we discuss some phenomenology resulting from this model. 
In Section 5 we present our conclusions and final remarks.

\section{Exotic Instantons and Quivers }

In this section, we briefly review D-brane instantons and unoriented quiver theories.

\subsection{Instantons}
In 4-dimensional gauge theory, instantons are point-like configurations, 
that extremize the Euclidean action for a given topological charge. 
In string theory, instantons admit a simple geometric interpretation: they are special Euclidean branes wrapping some (internal) cycle. In theories with (unoriented) open strings,  these are Euclidean D-branes (E-branes) that can intersect the `physical' D-branes\footnote{For a pedagogical review see {\it e.g.} \cite{Bianchi:2009ij, Bianchi:2012ud}.}. 
In (un-)oriented type IIA, gauge instantons can be classified as Euclidean D2 (E2) branes wrapping the same 3-cycle as a stack of ``physical'' D6-branes.
In (un-)oriented IIB, instantons are E(-1) or E3 wrapping the same holomorphic divisor as a stack of ``physical'' D7-branes. 
In type I, instantons are E5 branes in the internal space,  with the same magnetization as the D9, wrapping the entire $CY_{3}$.  

\subsection{Quivers}
The effective low energy description of the dynamics of D-branes 
at Calabi-Yau singularities is captured by a quiver field theory. 
Usually, the (supersymmetric) quiver conventions are the following: 
the standard D-brane stacks are `circle'
nodes,  the super-fields in the bi-fundamental representations of two D-brane stacks 
are oriented lines connecting the nodes, usually termed arrows whence the name `quiver', `triangle' nodes are Euclidean D-branes (instantons),
 grassmanian moduli or modulini are dashed lines connecting triangles and  circles.
 Square nodes represent flavour branes, {\it i.e.} branes wrapping non-compact cycles so much so that 
 $g^{2}_{YM, Dp} = g_s (\alpha^\prime)^{p+1/2}/V_{p+1} \rightarrow 0$.
 These simple rules allow one to subsume the system of D-branes' stacks and open strings 
 with a simple diagram. 
  In this notation, perturbative interaction terms involving  
 the matter super-fields correspond to closed oriented polygons, starting with triangles. On the other hand, interactions between standard super-fields and modulini
also correspond to closed oriented polygons involving solid and dashed lines\footnote{One should keep in mind that exotic instantons are brane wrapping empty nodes of the quiver. Their interactions are thus coded in the quiver or dimer.}.


\section{An Unoriented Quiver for a Pati-Salam model}

\begin{figure}[t]
\centerline{ \includegraphics [height=6cm,width=0.5 \columnwidth]{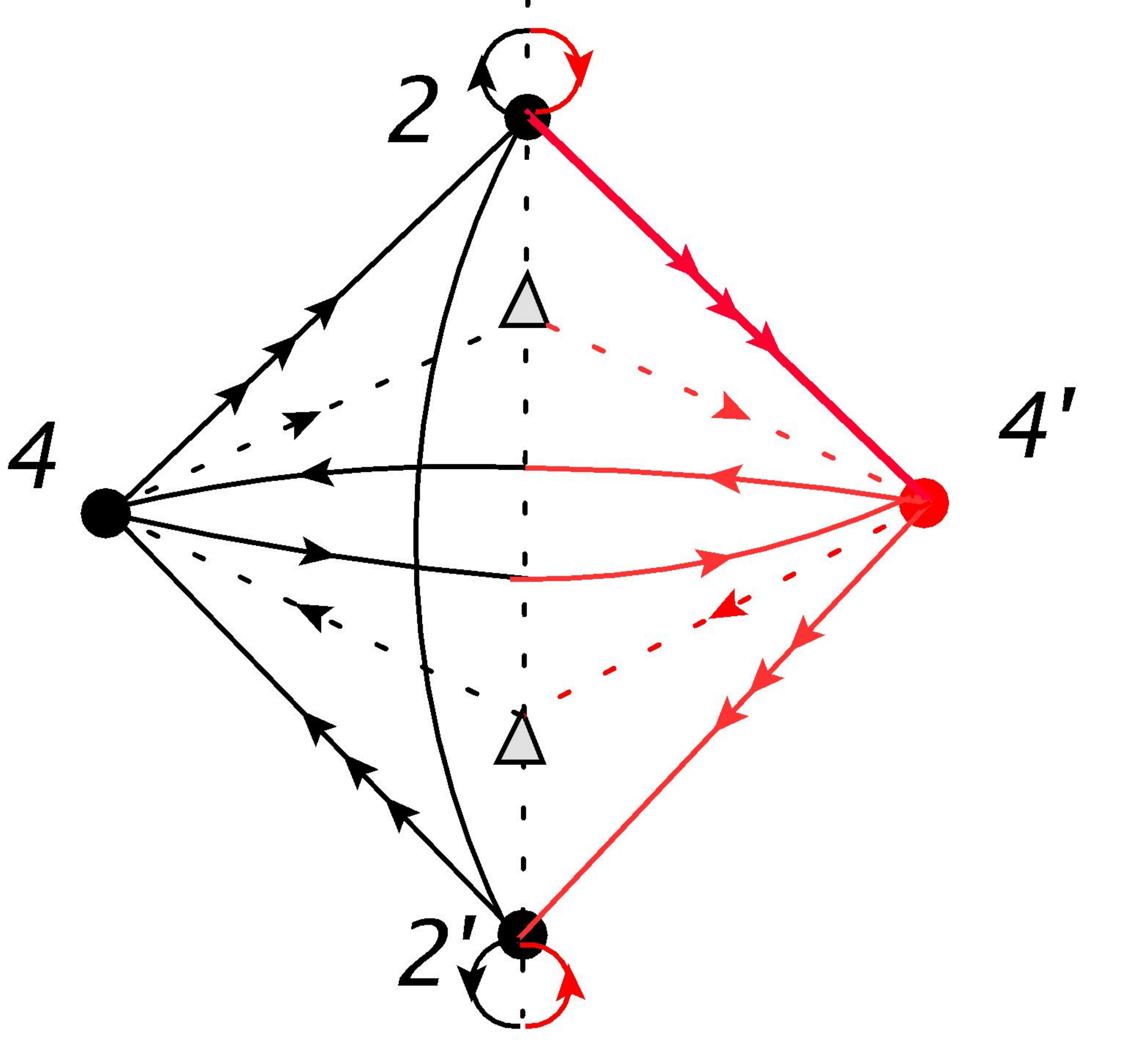}}
\vspace*{-1ex}
\caption{Unoriented quiver for a Pati-Salam-like model $U(4)\times Sp(2)_{L}\times Sp(2)_{R}$. The circles labeled by $4,4',2,2'$ represent $U(4),U'(4),Sp(2)_{L}, Sp(2)_{R}$ stacks respectively. 
An $\Omega^+$-plane identifies the $U(4)$ stack with its mirror image, 
$Sp(2)_{L,R}$ are stacks of two D6-branes laying exactly on top of the $\Omega$-plane. The symmetric representations $\Delta(1,1,10)$ and $\Delta^c(1,1, 10^*)$ appear in between the two stacks $4$ and $4'$. 
The triangles represent two possible $E2$-brane $O(1)$ instantons, laying on top of the $\Omega$-plane. 
 }
\label{plot}   
\end{figure}

\begin{figure}[t]
\centerline{ \includegraphics [height=6cm,width=0.5 \columnwidth]{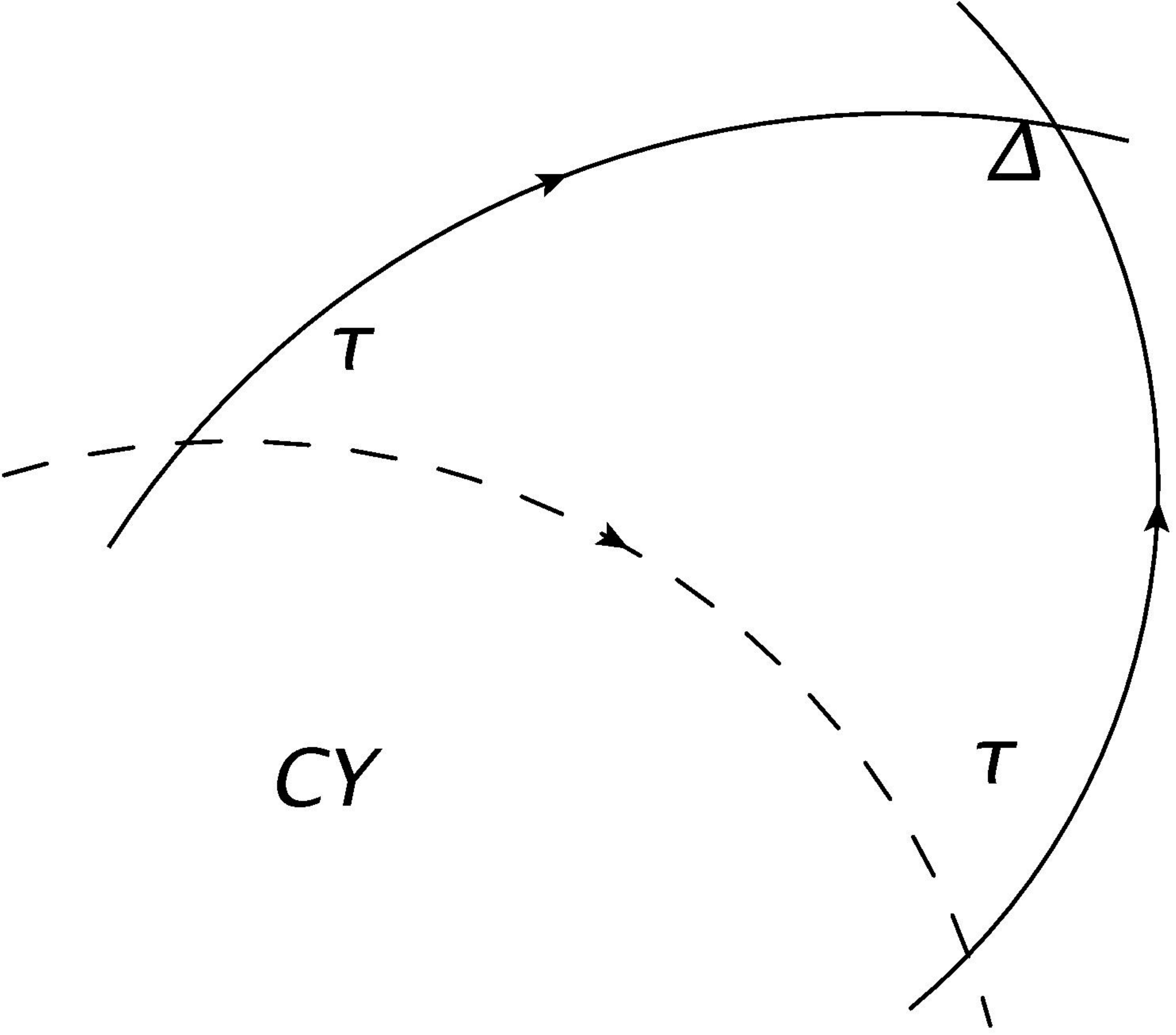}}
\vspace*{-1ex}
\caption{Amplitude in IIA (un)oriented string theory. $\Delta_{ab}$ sextets are excitations of strings attached between two intersecting $D6$-branes, represented in figure as black lines. The fermionic moduli (or modulini) result from strings localised at the intersection of one $D6$-brane and an exotic instanton Euclidean $D2$-brane (or $E2$ brane). They are represented in figure as dashed lines.  This system is embedded in some Calabi-Yau compactification $CY_{3}$. In particular $D6$-branes are wrapping 3-cycles on $CY_{3}$ and $E2$ are wrapping a different 3-cycle.   }
\label{plot}   
\end{figure}
 
 In this section, we construct  a simple quiver for a Pati-Salam model
 inducing a Majorana mass for the neutron.
 We propose a simple quiver in Fig.~2, leading to 
 an $\mathcal{N}=1$ susy Pati-Salam like model, with all the necessary 
 fields and Yukawa's for a spontaneous symmetry breaking pattern
 to the Standard Model and for the generation of a 
Neutron Majorana mass from Exotic Instantons. 
 
The gauge group is $U(4)\times Sp(2)\times Sp(2)$:
  $U(4)$ is generated by a stacks of 4 D-branes and their images
$U'(4)$ under an $\Omega^{+}$-plane;
  $Sp(2)_{L,R}$ are generated by two stacks of two D-branes each
on top of the $\Omega^{+}$-plane.
  We also consider Exotic $O(1)$ Instantons corresponding to 
$E2$-branes on top of the $\Omega^{+}$-plane.

In particular, the three generations of Left and Right  fundamental representations $F_{L,R}$, 
 containing quarks and leptons, are reproduced as excitations 
 of open strings attached to the $U(4)$-stack and the Left or Right $Sp(2)_{L,R}$
 stacks, respectively. We also get $\Delta=(1,1,10)$ and its conjugate from open strings attached 
 to the $U(4)$-stack and its mirror image $U(4)'$-stack. 
 $\phi_{RR}=(3,1,1)$ and $\phi_{LL}=(1,3,1)$ correspond to
strings with both end-points attached on the $Sp(2)_{R,L}$ respectively.
 Higgs fields $h_{LR}=(2,2,1)$ are between $Sp(2)_{L}$ and $Sp(2)_{R}$.
 It is amusing to observe that a two-generation model of the same kind would result in unoriented Type IIB from `fractional' D3-branes at a ${\bf C}^3/Z_4$ orbifold singularity. 
  
The perturbative super-potential that we obtain from the quiver reads 
 \be{Lang}
 \mathcal{W}_{eff}\sim y_{1}h_{LR} F_{L}F_{R}+\frac{1}{M_{1}}F_{L}\phi_{LL}F_{L}\Delta+\frac{1}{M_{2}}F_{R}\phi_{RR}F_{R}\Delta^{c}+\frac{1}{M_{3}}h_{LR}\phi_{RR}h_{RL}\phi_{LL}+\mu h_{LR}h_{RL}
 \ee
 $$+m_{\Delta}\Delta\Delta^{c}+\frac{1}{4 M_{4}}(\Delta\Delta^{c})^{2}+\frac{1}{2}m_{L}\phi_{LL}^{2}+\frac{1}{2}m_{R}\phi_{RR}^{2}+\frac{1}{3!}a_{L}\phi_{LL}^{3}+\frac{1}{3!}a_{R}\phi_{RR}^{3}$$
 where $\Delta\equiv\Delta_{qq}$ and the mass scales $M_{1,2,3,4}$ depend on the particular global completion of the model: they could be near $M_{S}$ or at lower scales. 
In a T-dual Type IIB context, the mass terms $m_{\Delta}$ and $m_{L,R}$ can be generated
by RR-RR or NS-NS three-forms fluxes in the bulk:
$$m_{\Delta_{qq}}\sim\Gamma^{ijk}\langle\tau H^{(qq)}_{ijk}+iF^{(qq)}_{ijk}\rangle$$
$$m_{L,R}\sim\Gamma^{ijk}\langle\tau H^{(L,R)}_{ijk} + iF^{(L,R)}_{ijk}~\rangle$$
with $H$ RR-RR and $F$ NS-NS three-forms
and in general $H^{dd},F^{dd}\neq H^{uu},F^{uu}$ depend on the choice of fluxes through the relevant cycles wrapped by the D-branes\footnote{Mass deformed quivers and dimers have been recently investigated in \cite{Bianchi:2014qma}.}.

On the other hand, the non-perturbative superpotential term:
\be{W}
\mathcal{W}_{E2} =\frac{1}{\mathcal{M}_{0}}\epsilon^{ijkl}\epsilon^{i'j'k'l'}\Delta^{c}_{ii'}\Delta^{c}_{jj'}\Delta^{c}_{kk'}\Delta^{c}_{ll'}
\ee
can be generated by an $E2$-brane instanton that intersect twice with the 
$U(4)$ stack of $D6$-branes so as to produce a four-$\Delta^c$ (as well as a four-$\Delta$) interaction.
The fermionic modulini $\tau^i_\alpha$, with $i=1,..., 4$ and 
$\alpha, \beta =1,2$ interact with the super-fields $\Delta$'s via
\be{Exotic}
\mathcal{L}_{E2-D6-D6}\sim \tau^{i}_{\alpha}\Delta^{c}_{(ij)}\tau^{j}_{\beta}\epsilon^{\alpha \beta}+\tau_{i\alpha'}^{c}\Delta^{(ij)}\tau_{j\beta'}^{c}\epsilon^{\alpha'\beta'}
\ee
These interactions are induced by mixed disk amplitudes as the one in Fig.~4, that emerge at the intersections 
between two $D6$-brane stacks and one $E2$ instanton\footnote{For similar calculations in related contexts, see \cite{Blu1,Ibanez1,Ibanez2, BiaKir, BiaFucMor, Addazi:2014ila}.}. In our case the two 
$D6$-branes are actually the $D6$-branes of the $U(4)$ stack and their images. Integrating out the fermionic 
modulini produces two $\epsilon^{ijkl}$ so that 
\be{W22}
\mathcal{W}_{E2} = \frac{1}{\mathcal{M}_{0}}\int d^{8}\tau e^{\mathcal{L}_{E2-D6-D6}} = 
\frac{1}{\mathcal{M}_{0}}\epsilon^{ijkl}\epsilon^{i'j'k'l'}\Delta^{c}_{ii'}\Delta^{c}_{jj'}\Delta^{c}_{kk'}\Delta^{c}_{ll'}
\ee
The suppression scale $\mathcal{M}_{0}$ is related to the string scale $M_{S}$ by 
$\mathcal{M}_{0}\sim M_{S}e^{+S_{E2}}$, where $S_{E2}$ depends on the closed string moduli that parametrize the (complex) size of the 3-cycles, wrapped by the $E2$-instantons\footnote{In general, the calculations could be much more complicated, in the
presence of bulk fluxes, that can also induce soft susy breaking mass terms for the 
susy partners. For example gaugino mass terms with $M_{\lambda}~\sim~\Omega_{0,3}^{ijk} \langle \tau H_{ijk}+iF_{ijk} \rangle$ can be generated in Type IIB contexts
by internal 3-form fluxes. In the presence of fluxes, one has to verify that physical branes  
and instantons are not lifted, {\it i.e.} the cycles they wrap and their intersections
are not eliminated. 
With the introduction of bulk fluxes, one also has to consider the back-reactions  on the exotic instantons, that could change then number of zero modes. 
This could modify our present analysis.}. 

After symmetry breaking $SU(4) \rightarrow SU(3) \times U(1)_{B-L}$,
 ${\bf 10_{+}}\rightarrow ({\bf 6}_{+2/3},{\bf 3}_{-2/3},{\bf 1}_{-2})$.
 We denote by $\Delta_{6}$ the `diquark' super-field in the ${\bf 6}$, $T_{3}$ in the triplet ${\bf 3}$
 and $S$ the singlet ${\bf 1}$ and 
 find
 \be{breaking}
\frac{1}{M_{2}}\phi^{c}_{\dot{\alpha}\dot{\beta}}F_{i}^{c^{\dot{\alpha}}}F_{j}^{c^{\dot{\beta}}}\Delta^{c^{(ij)}} \rightarrow \frac{1}{M_{2}}\phi_{\dot{\alpha}\dot{\beta}}\left[Q^{c^{\dot{\alpha}}}_{i}Q^{c^{\dot{\beta}}}_{j}\Delta_{6}^{c^{ij}}+ 2~Q_{i}^{c^{\dot{\alpha}}}L^{c^{\dot{\beta}}}T_{3}^{c^{i}}+L^{c^{\dot{\alpha}}}L^{c^{\dot{\beta}}}S^{c} \right]
 \ee
  \be{Lmasse}
 m_{\Delta}\Delta\Delta^{c}+\frac{1}{4M_{4}}(\Delta\Delta^{c})^{2}
 \rightarrow m_{\Delta}(\Delta_{6}\Delta_{6}^{c}+TT^{c}+SS^{c})+\frac{1}{4M_{4}}(\Delta_{6}\Delta_{6}^{c}+TT^{c}+SS^{c})^{2}
 \ee
\be{delta}
\frac{1}{\mathcal{M}_{0}}\epsilon_{ijkl}\epsilon_{i'j'k'l'}\Delta^{c^{ii'}} \Delta^{c^{jj'}} \Delta^{c^{kk'}} \Delta^{c^{ll'}}
\ee
$$ \rightarrow \frac{1}{\mathcal{M}_{0}}\left[4~\epsilon^{SU(3)}_{ijk}\epsilon^{SU(3)}_{i'j'k'}\Delta_{6}^{c^{ii'}}\Delta_{6}^{c^{jj'}}\Delta_{6}^{c^{kk'}}S+6~\epsilon^{SU(3)}_{ijk}\epsilon^{SU(3)}_{i'j'k'}\Delta_{6}^{c^{ii'}}\Delta_{6}^{c^{jj'}}T_{3}^{c^{k}}T_{3}^{c^{k'}} \right] $$ 
The complete super-potential after symmetry breaking $SU(4) \rightarrow SU(3) \times U(1)_{B-L}$
 reads 
 \be{Lagrf}
 \mathcal{W}\sim y_{1}h_{\alpha \dot{\alpha}}Q^{i\alpha}Q^{c^{i\dot{\alpha}}}+y_{1}h_{\alpha \dot{\alpha}}L^{\alpha}L^{c^{\dot{\alpha}}}+\frac{1}{M_{1}}\phi_{\alpha \beta}(Q^{i\alpha}Q^{j\beta}\Delta_{ij}+Q^{i\alpha}L^{\beta}T_{3i}+L^{\alpha}L^{\beta}S)
 \ee
 $$+\frac{1}{M_{2}}\phi_{\alpha \beta}^{c}(Q^{c^{i\alpha}}Q^{c^{j\beta}}\Delta^{c}_{ij}+Q^{c^{i\alpha}}L^{c^{\beta}}T_{3i}^{c}+L^{c^{\alpha}}L^{c^{\beta}}S^{c})+\mu h_{\alpha \dot{\alpha}} h^{\alpha \dot{\alpha}}$$
 $$+\frac{1}{M_{3}}h^{c}\phi^{c}h\phi+m_{\Delta}(\Delta_{6}\Delta_{6}^{c}+TT^{c}+SS^{c})+\frac{1}{4M_{4}}(\Delta_{6}\Delta_{6}^{c}+TT^{c}+SS^{c})^{2}+W(\phi_{L,R})$$
 $$+\frac{1}{\mathcal{M}_{0}}\left[4~\epsilon^{SU(3)}_{ijk}\epsilon^{SU(3)}_{i'j'k'}\Delta_{6}^{c^{ii'}}\Delta_{6}^{c^{jj'}}\Delta_{6}^{c^{kk'}}S+6~\epsilon^{SU(3)}_{ijk}\epsilon^{SU(3)}_{i'j'k'}\Delta_{6}^{c^{ii'}}\Delta_{6}^{c^{jj'}}T_{3}^{c^{k}}T_{3}^{c^{k'}} + (\Delta^c,T^c,S^c \rightarrow \Delta, T, S) \right]$$
where $W(\phi_{L,R})=m_{L,R}\phi_{L,R}^{2}+a_{L,R}\phi_{L,R}^{3}/3$.
 
When $\langle\phi_{RR}\rangle=v_{R}$ and $\langle\phi_{LL}\rangle=0$, Left-Right symmetry is spontaneously broken. 
When $\langle S\rangle=v_{B-L}$, $SU(4)$ and its subgroup $U(1)_{B-L}$ are spontaneously broken. 
A Majorana mass term for the Neutrino is generated\footnote{Dirac masses are generated via 
Yukawa couplings when $h_{LR}$ gets a VEV.} as
$m_{N} \sim v_{R}v_{B-L}/M_{2}$. 
For example, $m_{N}\simeq 10^{12}\, \rm GeV$
can be obtained if $v_{R}\simeq M_{2}$
and $v_{B-L}\simeq  10^{12}\, \rm GeV$.
In this model the generation of a neutrino Majorana
mass is connected to the generation of a Neutron Majorana mass.

In fact, when $S$ takes an expectation value, 
a cubic interaction term 
$$(v_{B-L}/\mathcal{M}_{0})\epsilon^{SU(3)}_{ijk}\epsilon^{SU(3)}_{i'j'k'}\Delta_{6}^{c^{ii'}}\Delta_{6}^{c^{jj'}}\Delta_{6}^{c^{kk'}}$$
 is generated. 
 In the next section, we will discuss 
 the consequences, for Neutron-Antineutron physics
 and for LHC phenomenology, in more in details.

\section{Neutron-Antineutron oscillation through color diquark sextets}

In a susy PS-like model $SU(4)\times SU(2)_{R}\times SU(2)_{L}$,
one can construct a diagram like the one in Fig.~1 for Neutron-Antineutron oscillation, 
through the `exotic' interaction $\Delta_{10}^{c}\Delta_{10}^{c}\Delta_{10}^{c}\Delta_{10}^{c}/\mathcal{M}_{0}$,
containing 
\be{maininside}
\mathcal{W}_{\Delta B=2}=\frac{1}{\mathcal{M}_{0}}\epsilon^{u^{c}d^{c}d^{c}\nu^{c}}\epsilon^{u'^{c}d'^{c}d'^{c}\nu'^{c}}\Delta^{c}_{u^{c}u'^{c}} \Delta^{c}_{d^{c}d'^{c}} \Delta^{c}_{d^{c}d'^{c}}S^{c}_{\nu^{c}\nu'^{c}} 
\ee
(with $S^{c}_{\nu^{c}\nu^{c}}\equiv \Delta^{c}_{\nu^{c}\nu^{c}}$). 
The operator (\ref{maininside}) induces a neutron-antineutron transition depicted in Fig.~2, 
as a result of the super-potential term 
$\tilde{f}_{11}v_{R}Q^{c}Q^{c} \Delta^{c}/M_{2}$, whose components include $f_{11}\Delta^{c}_{u^{c}u^{c}}u^{c}u^{c}$ and $f_{11}\Delta^{c}_{d^{c}d^{c}}d^{c}d^{c}$, with $f_{11}=\tilde{f}_{11}v_{R}/M_{2}$,
and $\tilde{f}_{11}$ Yukawa couplings. 

The process in Fig.~1-(b) produces an effective operator $G_{n-\bar{n}}(udd)^{2}$ with
\be{Gnnbar2}
G_{n-\bar{n}}\simeq \frac{g_{3}^{2}}{16\pi}\frac{f_{11}^{2}v_{BL}}{M_{\Delta^{c}_{u^{c}u^{c}}}^{2}M_{\Delta^{c}_{d^{c}d^{c}}}^{2}M_{SUSY}\mathcal{M}_{0}}
\ee

We can now discuss different choices of the parameters leading to very different branches for phenomenology. 
The motivation of such a variety of possibilities is related to the fact that in (\ref{Gnnbar2})
one can produce a scale of $300-1000\, \rm TeV$, testable in the next generation of experiments, with different choices of the other parameters. 

The cases with $\mathcal{M}_{0}\simeq 10^{19}\,\rm GeV$ and $\mathcal{M}_{0}\simeq 10^{13}\,\rm GeV$ are equivalent to the GUT $SO(10)$ inspired scenario, discussed in \cite{Md} in Fig.~1-(b).
In these cases $M_{\Delta^{c}_{u^{c}u^{c}}}\sim 1\, \rm TeV$. 
Both cases are well compatible with the mechanism proposed in the previous section.
In fact a scenario in which $\mathcal{M}_{0}\simeq M_{S}$ can be envisaged, if $e^{S_{E_{2}}}\sim 1$ {\it i.e.} small 3-cycles wrapped by $E_{2}$ in $CY$.
{\it A priori}, the string scale can be considered as a free parameter, it can be as  high as $10^{19}\, \rm TeV$ as low as a few $\rm TeV$'s.   
For instance, if $M_{S}=1\div 10\, \rm TeV$, the hierarchy problem of the Higgs mass is automatically solved, 
and $\mathcal{M}_{0}$ can be as high as $10^{13}\, \rm GeV$ (or more) if $e^{S_{E_{2}}}=10^{10}$ {\it i.e.} for an $E_{2}$  wrapping a large 3-cycle in the $CY$. 

In TeV-scale gravity scenari, one can also consider an alternative scenario in which $\mathcal{M}_{0}\simeq M_{S}\simeq 10\, \rm TeV$, 
with $e^{S_{E2}}\sim 1$.  Compatibly with $n-\bar{n}$ limits, 
the four-sextets' interaction (\ref{W}) is suppressed at much lower scales, 
with  $\mathcal{M}_{0} \sim 10\, \rm TeV$. 
In this case, the masses of the sextets have to be much higher than $1\, \rm TeV$, eluding completely a direct observation at LHC and in FCNCs. 
On the other hand, a post sphaleron baryogenesis scenario as the one proposed in 
\cite{Md} remains possible. 

So, combined observations from the next generation of experiments on neutron-antineutron oscillations, 
LHC, FCNC physics, neutrino physics can provide precious informations not only on a possible PS extension of the SM and its region of the parameters, 
but also on the dynamical scale generated by the Exotic instantons, and as a consequence on the geometric structure 
of the Calabi-Yau compactifications, in particular the 3-cycle wrapped by the Exotic Instanton! 
This is a fundamental information for realistic model building in string phenomenology. 
 Exotic instantons could be {\it portals} from low energy physics to the quantum gravity scale!
In the next section, we will briefly discuss connections with dark matter and the hierarchy problem.
Note that in our scenario, $\langle S\rangle\neq \langle\phi_{RR}\rangle$ in general. This is an important difference with respect 
to the Babu-Mohapatra model cited above: $n-\bar{n}$ oscillation time of order $10^{10}\, \rm s$ is compatible 
with a Left-Right symmetry restoration at the TeV scale, with intriguing implications for LHC.
A recent anomaly in $pp\rightarrow l_{1}l_{2}jj$ with significance near $3\sigma$, compatible with Left-Right symmetry,  
was seen by CMS \cite{CMS}. In a Left-Right model, this is interpreted as 
sequential $W_{R}$ and $N_{R}$ production as \cite{CMS1,CMS2}
$$pp\rightarrow W_{R}\rightarrow l_{1}N_{R}\rightarrow l_{1}l_{2}W^{*}_{R}\rightarrow l_{1}l_{2}jj$$
However, this interpretation requires $g_{R}(M_{{W}_{R}})\simeq 0.6 g_{L}(M_{W_{R}})$,
with $1.8\, \rm TeV<M_{W_{R}}<2.4\, \rm TeV$. 
Curiously, this situation is not compatible with a D-parity preserving $SO(10)$ GUT scenario.
For P-S models emerging from $SO(10)\rightarrow SU(4)\times SU(2)_{L}\times SU(2)_{R}$
D-parity is a symmetry. It is the external automorphisms that exchanges the two $SU(2)$ groups,
$SU(2)_{L}\leftrightarrow SU(2)_{R}$, and at the same time acts by conjugation on $SU(4)$ representations. More explicitly $D_P = \Gamma_7 \gamma_5$ is a symmetry when $g_{L}=g_{R}$. At the unification scale $g_{L}=g_{R}$,  the two coupling constants have the same running if the field content is LR symmetric.  For a LR interpretation of CMS anomaly one needs $g_{L}\neq g_{R}$, compatibly with a PS model not emerging from $SO(10)$ without D-parity altogether or  at least an $SO(10)$ model where D-parity is broken at a high scale. 
On the other hand, D-parity is not a symmetry in string-inspired models.
The gauge couplings depend on the size of the internal cycle wrapped by the D-branes. In orbifolds or Calabi-Yau singularities, one can tune the blow-up modes so that different cycle have the same size {\it e.g.} vanishing, but generically this is not the case and one can start with $g_{L}\neq g_{R}$ already at the string scale.

$pp\rightarrow l_{1}l_{2}jj$ is not the only peculiar channel suggested by our model for LHC, 
and we will return to other ones in the next section. 
We conclude this section with another observation. 
A scenario in which $S$ doesn't take an expectation value at all can be envisaged in our model. 
In this case, $U(1)_{B-L}$ is not spontaneously broken by $S$.
However, as mentioned above, exotic intantons can dynamically break $U(1)_{B-L}$.
 For instance, a Majorana mass matrix for RH neutrini
can be generated by exotic instantons rather than by $S$, as cited above. 
In this case, $S$ could also be a light particle, if a residual discrete symmetry 
of $U(1)_{B-L}$ stabilizes it. In other words, $S$ can behave as 
a Majoron, but it is not exactly a Majoron \cite{Majoron}. We can call it an {\it exoticon}.
We suppose that the exoticon interacts with the three color sextets with a coupling $\mu_{S}$.
So, in this case, we have to replace $v_{B-L}$ with $\mu_{S}$ in (\ref{Gnnbar2}).
The exoticon carries $B-L=-2$, so $n\rightarrow \bar{n}S$
does not violate $B-L$.
We also note another important difference with respect to Majorons:
the Majoron mass $m_{\phi}=y_{L}v_{L}$,
with $v_{L}$ vev of a global $U(1)_{L}$ (and $y_{L}$ coupling),
 is related to its interaction with neutrini,
 as $g_{\phi\nu\nu}=m_{\nu}/v_{L}$;
 such a relation, in general, is not satisfied by exoticons. 
A massive exoticon cannot be emitted in a $n-\bar{n}$ transition, in the vacuum: CPT symmetry protects
neutron by transitions $n\rightarrow \bar{n}+S$, i.e $m_{n}=m_{\bar{n}}$. 
However, in a nuclear environment, such a transition is allowed!
Such a transition is followed by annihilation of the antineutron with another neutron in 
the nuclear environment, 
as $(Z,A) \rightarrow (Z,A-2)+3\pi$.
We can roughly estimate the corresponding decay width
to be $\Gamma \simeq (\delta m/\mu_{S})^{2}\langle\Delta \mathcal{E}\rangle$,
where $\langle\Delta \mathcal{E}\rangle\simeq 10\div 100\, \rm MeV$ is the 
average energy in the nuclear environment. 
Limits on $n-\bar{n}$ oscillation in the nuclei are $\Gamma_{n\bar{n}}^{-1}\sim 10^{-32}\, \rm yr$ \cite{PDG},
corresponding to $\mu_{S}> 10^{30}\delta m \simeq \rm keV$.
Another spectacular signature of an exoticon could be
a nucleon-nucleon disappearances as $nn\rightarrow S$, $\Delta B=2$.
This could be detected as a nuclear transition 
$(Z,A)\rightarrow (Z,A-2)+\rm missing\, \rm energy$.
We can easily estimate the rate of such a transition as
$\Gamma \sim \kappa_{np} (\mathcal{M}_{n\bar{n}})^{-3} m_{N}^{14}G_{n-\bar{n}}^{2} \, \rm GeV$,
where $\kappa_{np}\sim 10^{-6}$ approximately accounts for the hadronic 
non-perturbative correction. Such an estimate leads us to conclude that
such a process is very suppressed, roughly as $10^{40}\, \rm yr$. 
Finally, an exoticon can also be  
detected in a neutrinoless double-beta-decay, as a Majoron.
However, there are several important differences with respect to the Majoron:
a $0\nu 2 \beta+S$ does not violate lepton number, it is an {\it apparent violation}.
For a $0\nu2\beta+S$ process, limits 
on the exoticon production imply 
$(m_{\nu}/\mu_{S})<10^{-5}$ \cite{GERDA,EXO}
 that corresponds to a bound $\mu_{S}>10\, \rm keV$. 
  Limits from supernovae cooling 
 processes $\nu \rightarrow \nu^{c}S$ or $\nu\nu \rightarrow SS$ 
 are competitive  (for electronic neutrini $m_{\nu}/\mu_{S}<10^{-5}$) \cite{Supernovae}.

\section{Other Comments on Phenomenology, Dark Matter and hierarchy problem}
 In Minimal non-susy Left-Right models,  identifying a candidate for cold dark matter is difficult. 
On the other hand, our model automatically suggests several candidates for cold dark matter.
In fact, our model predicts the presence of neutralini and stuckelberg axini (or stuckelini),
mixing with each other, as in \cite{axini1,axini2}. They are good candidate for WIMP dark matter.
On the other hand, $Z'$ corresponding to anomalous $U(1)$ can get a mass from Stuckelberg mechanism, 
that is not necessary of the order of string scale $M_{S}$. 
In particular, a scenario in which $M_{S}\simeq 10 \div 10^{3}\, \rm TeV$ can be envisaged, 
alleviating the hierarchy problem of the Higgs mass.
In this case, $m_{Z'}\simeq 10^{-4}M_{S}\div M_{S}\simeq 1\div 10\, \rm TeV$, 
with Generalized Chern Simon terms inducing peculiar decays as
$Z' \rightarrow Z\gamma$ \cite{212,213,214,215,Anastasopoulos:2006cz, 216,217,218, axini1,axini2,Stuck4}\footnote{We would like to stress that GCS terms 
generate UV divergent triangles that are cured by considering UV completions with KK states or string excitations.
For issues in scattering amplitudes and collider physics see \cite{Santini}, for recent discussion 
about string theory and causality see \cite{Maldacena,Veneziano}. For non-local (string inspired) quantum field theories see
\cite{Anumap1,B1,Addazi:2015dxa}}. Curiously, the presence of apparent flavor violations 
in $B\rightarrow llK$, detected in LHCb \cite{LHCb1,LHCb2,LHCb3,LHCb4} could be a hint in favor of a new $Z'$ \cite{LHCb5,LHCb6,LHCb7}.
In our model, we also expect extra decays $B \rightarrow l^{+}l^{-}l'^{+}l'^{-}K$ suppressed by the GCS couplings 
with respect to $B\rightarrow l^{+}l^{-}K$.  
Extra $Z'$ from anomalous symmetries are different with respect to $Z'_{R}$ ($Z$-boson of the $SU(2)_{R}$), 
and kinetic mixings $Z-Z'$ or $Z'-Z'_{R}$ can be envisaged. 
 $Z'_{R}$ can also interact with $Z,\gamma,Z'$ through G.C.S. A complete study of the
resulting cascade processes is beyond the purpose of this paper.
Concerning the hierarchy problem of the Higgs mass, our model is compatible with TeV-scale supersymmetry, 
but this model has more undetermined parameters than the MSSM, {\it i.e.} it could be more elusive and more difficult to constrain.

\section{Conclusions and remarks}

In this paper, we have shown how to generate a Majorana mass for the neutron, inducing neutron-antineutron transitions with $|\Delta B| = 2$
in the context of a Pati-Salam Left-Right model. 
Indeed exotic instantons can produce an effective interaction involving color diquark sextets,
leading to a Majorana mass term for the neutron. We would like to stress that in the present context no processes with $|\Delta B| = 1$ are allowed that could lead to fast proton decay.
We have discussed some possible phenomenological implications, in the main branches of the parameters space,
for LHC, FCNC, Dark Matter, $0\nu2\beta$-decay.
A unifying picture of Dark Matter, Hierarchy problem of the Higgs mass, Baryogenesis
and Neutrino mass emerges in a very simple unoriented quiver!
In this sense, the model elaborated here could represent a serious alternative to GUT inspired models. 

\vspace{1cm} 

{\large \bf Acknowledgments} 
\vspace{3mm}

We would like to thank R.~Mohapatra and A.~Pascal for reading the manuscript and suggesting insightful comments. We would also like to thank C.~Bachas, F.~Quevedo and G.~Villadoro for useful discussions.   A.~A. would like to thank Galileo Galilei Institute for Theoretical Physics for the hospitality, where this paper was prepared.


\end{document}